\documentclass[
 reprint,
 nobibnotes,
 amsmath,amssymb,
 aps,
 prb,
]{revtex4-2}

\usepackage{graphicx}% Include figure files
\usepackage{dcolumn}% Align table columns on decimal point
\usepackage{bm}% bold math
\usepackage{braket}
\usepackage{upgreek}

\usepackage{xcolor}

\usepackage[overlay,absolute]{textpos}

\newcommand\PlaceText[3]{
	\begin{textblock*}{10in}(#1,#2)
		#3
	\end{textblock*}
}

\begin{document}

\title{Optical transmitter tunable over a 65-nm wavelength range around 1550 nm for quantum key distribution}

\author{Benjamin Griffiths, Yuen San Lo, James F. Dynes, Robert I. Woodward, and Andrew J. Shields}

\affiliation{Toshiba Europe Ltd., Cambridge, UK}

\begin{abstract}
The ability to create phase-controlled pulses of light with wavelength tunability has applications spanning quantum and classical communications networks. Traditionally, optical transmitters are able to either produce phase-controlled pulses at a fixed wavelength or require a chain of bulky and expensive external modulators to convert wavelength tunable continuous wave light into optical pulses. One technology of great interest is quantum key distribution (QKD), a technology for generating perfectly random keys at remote nodes to ensure secure communications. Environments such as data centers, where the user needs change regularly, will require adaptability in the deployment of QKD to integrate into classical optical networks. Here we propose and demonstrate an alternative quantum transmitter design consisting of a multi-modal Fabry-Perot laser optically injection locked by a wavelength tunable laser. The transmitter is able to produce phase-controlled optical pulses at GHz speeds with a tunable wavelength range of $>$65~nm centered at 1550~nm. With this transmitter, we perform proof-of-principle QKD with secure bit rates of order Mb/s.
\end{abstract}

\maketitle

\PlaceText{12mm}{8mm}{Phys. Rev. Appl. \textbf{20}, 044040 (2023); https://doi.org/10.1103/PhysRevApplied.20.044040}

\section*{Introduction}

Optical transmitters within the telecommunications industry often utilize compact laser sources based upon semiconductor lasers as they are able to reliably create high quality optical pulses at high speeds and low cost \cite{ACKERMAN2002587,doi:https://doi.org/10.1002/3527600434.eap867}. Very high pulse signal-to-noise extinction ratio optical pulses can be achieved by gain-switching a semiconductor laser via the direct application of a radio frequency (RF) signal to the gain-medium. The modulation bandwidth of the semiconductor laser, and thus the rate at which optical pulses can reliably be achieved, is proportional to the confinement factor of the laser cavity and inversely proportional to the active volume \cite{8859279}. Often distributed feedback (DFB) lasers are used for optical transmitters where pulse repetition rate and extinction ratio are important due to their high pulse performance. Other laser designs can be used but often with a loss in pulse performance \cite{Semenenko:20}. DFB lasers are formed by etching reflectors into the gain medium, giving single-mode operation, as well as high confinement in a small volume. This results in excellent modulation bandwidth. However, by etching mirrors (with a fixed reflection profile) into the gain medium, it is difficult to tune the wavelength, other than a few nm by temperature. To achieve wavelength tunability, an external cavity is needed to introduce some tunable element, but this increases the laser size and reduces confinement – reducing the modulation bandwidth.

Wavelength tunability is becoming a widely desirable feature in telecommunications transmitters as network designs develop to integrate more flexibility into the network architecture \cite{Mroziewicz,duarte2017tunable,green2001progress}. A wavelength tunable (WT) laser is the obvious starting point and these are commercially available and easy to integrate into current telecommunications networks \cite{10.1117/12.543338}. However, as outlined above their external cavity design is incompatible with high bandwidth gain-switching techniques \cite{duarte2017tunable}. One simple approach to mitigate this would be to place an external intensity modulator after the output of the WT laser to carve pulses from a continuous wave (CW) source \cite{winzer2004chirped}. However, to achieve a high pulse signal-to-noise extinction ratio of a similar level to that achieved through gain-switching with such a system, a chain of modulators, each with their own associated drive electronics, would be required which is expensive and bulky \cite{binh2006lithium}.

An application where high speed optical pulses is required is quantum key distribution (QKD). QKD is an information-theoretic secure method of communication through information encoding into conjugate basis states of quantum particles \cite{BB84,RevModPhys.74.145}. Fiber based QKD systems generally utilize photons at wavelengths in the communications C band (1530 nm--1565 nm) to encode the quantum information due to the reduced level of optical losses compared to other wavelengths travelling down a fiber \cite{QKDmultiplex,8595725}.

Integration with classical communications networks requires flexibility in choosing the wavelength within this band to send quantum information due to significant demand from classical devices which can disturb fragile quantum states via Raman scattering leading to higher error rates \cite{Aleksic:15,Mao:18,ToudehFallah2022PavingTW}. This is particularly important in large scale commercial networks where many network operators will be unable to change the wavelength of their classical traffic making flexibility for QKD a key requirement. Flexibility in wavelength of QKD deployment also opens the door to dynamically reconfigurable QKD where the ability to change the quantum and service channels based upon bandwidth availability will enable integration of QKD with emerging next-generation software-defined networks \cite{Hugues-Salas:18,Cao:19}.

\begin{figure*}[t]
\centering\includegraphics[width=\textwidth]{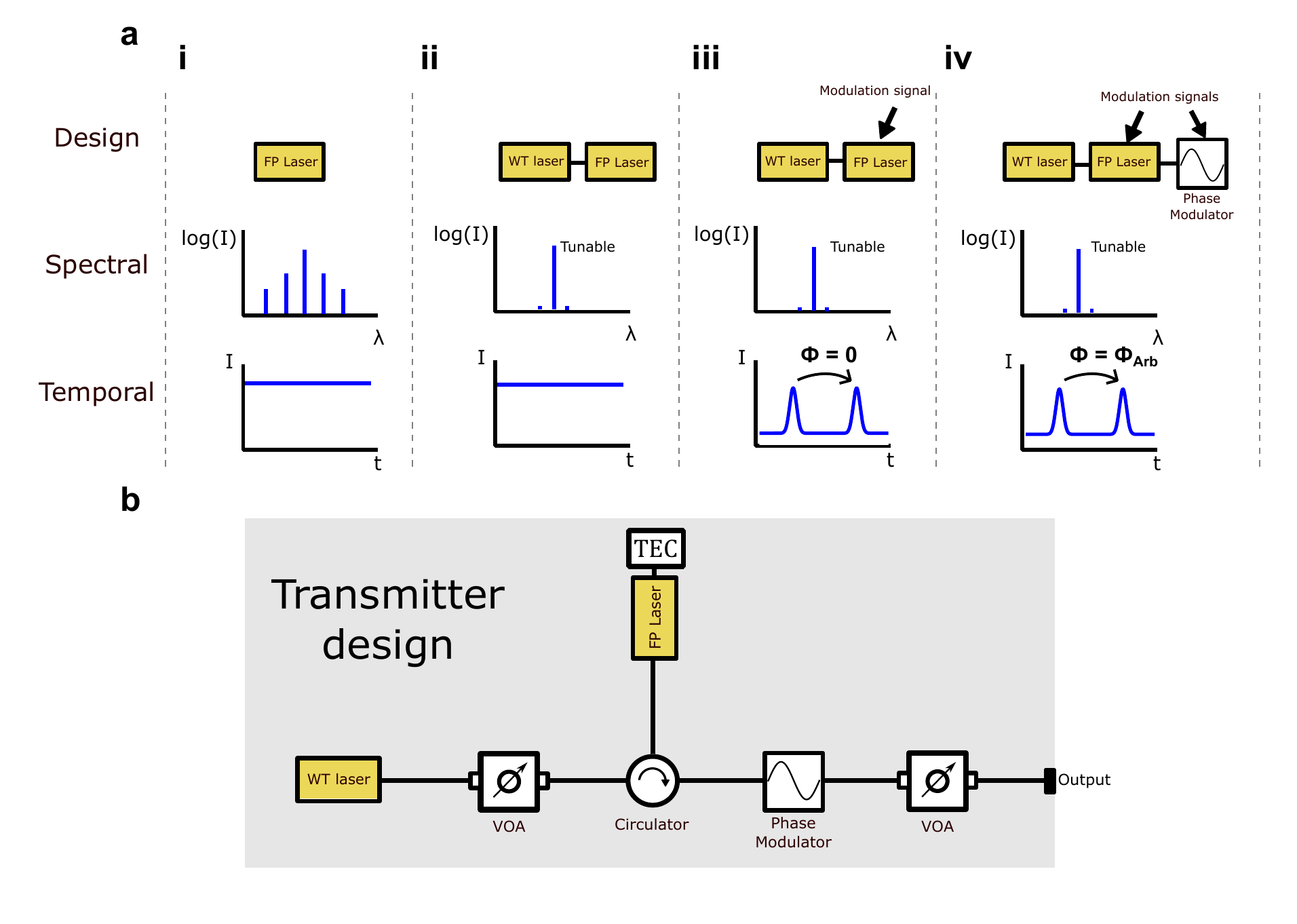}
\caption{\textbf{a} A schematic showing the concept of how to achieve wavelength tunable single mode pulses with phase control. \textbf{i} The direct output of a FP laser diode showing multi-modal spectral behaviour and a CW output. \textbf{ii} Injection locking of the FP laser with a WT laser enables a tunable single mode output operating in CW. \textbf{iii} Gain switching the FP laser enables wavelength tunable coherent pulses. \textbf{iv} Passing the pulses through a phase modulator enables phase-controlled optical pulses with a tunable wavelength. \textbf{b} The physical design of the transmitter showing the optical path used to injection lock the FP laser to the WT laser. A variable optical attenuator (VOA) is used to control the power of injection and the output flux. A thermoelectric coupler (TEC) is used to fine tune the mode positions of the FP laser to maximise the mode overlap with that of the injected light.}
\label{Concept}
\end{figure*}

Today's laser based QKD systems are generally restricted to a very small wavelength range due to the single-mode operation of the lasers that are used for fast pulse repetition speeds \cite{759384,Semenenko:19}. Wavelength tuning of these devices is typically restricted to a few nanometres due to the internal dynamics of the lasers used \cite{1303579}. Switching to a new wavelength therefore requires significant modification of the QKD hardware such as changing the laser source. Here we present a new design for a wavelength tunable quantum transmitter which is able to operate on demand at wavelengths ranging from 1520 nm to 1585 nm. The optical pulses are produced through the use of a gain-switched multi-modal Fabry-Perot (FP) laser which is optically injection locked (OIL) by an external cavity wavelength tunable (WT) laser source. Attenuating the output light to the single photon level, we use the transmitter alongside a standard QKD receiver as a wavelength tunable point-to-point QKD system. Information is encoded into time and phase bits operating a 1 GHz symbol clock-rate.

\section*{Transmitter design}

To achieve the wavelength tunable pulses of light with phase control, an injection seeding design was used, splitting the wavelength tunability and pulsing behaviour into two separate lasers \cite{chipmodfree,8859279,Barry2001OpticalPG,4838882,Semenenko:19}. An external cavity tunable semiconductor laser with wavelength tunability between 1500 nm and 1600 nm and linewidth of 100 kHz was used in CW operation to select the chosen wavelength. The external-cavity design enables wide tunability, but limits the high-speed performance of the laser. We circumvent this limitation by using the light to injection lock a single mode of a multi-modal FP laser, where the FP laser geometry facilitates significantly improved pulsed and high-bandwidth modulation performance. The FP laser had a modulation bandwidth of 10 GHz and modulation was applied by directly coupling an RF signal to the gain-medium. The phase of FP laser outputted pulses is set by the phase of the injected light such that all pulses are coherent with each other. Applying phase modulation through use of a high bandwidth external lithium niobate (LiNbO$_{3}$) phase modulator then allows an arbitrary phase to be added to any pulse. Such lasers and optical components are ubiquitous in the telecommunications industry, hence practical systems based on the ideas in this paper can be built with off-the-shelf telecommunications components. Indeed, similar designs have been used before in classical communications networks to produce high speed wavelength tunable optical pulses, however in these cases the phase relationship of the pulses is not carefully controlled \cite{Barry2001OpticalPG, lau2007frequency, 4838882, pascual2017photonic}. Therefore, we build upon this work by considering new encoding approaches and carefully exploring the related laser physics to tailor the transmitter to the needs of QKD applications.

The FP laser was chosen as it has a broad gain profile that can support optical modes across a large range of wavelengths. FIG. \ref{Concept}a shows the concept that was used for producing wavelength tunable phase-controlled pulses. The FP laser has a multi-modal structure across tens of nanometres of range (FIG. \ref{Concept}a.i). The WT laser was then injected into the FP laser such that it overlapped with one of the modes (FIG. \ref{Concept}a.ii). Tuning the temperature of the FP laser with a thermoelectric coupler (TEC) permitted fine tuning of the FP laser mode positions allowing maximisation of the overlap between the FP laser mode and the injected mode, as well as giving access to a continuum of wavelengths. The input light dominated the stimulated light emission within the FP laser cavity causing significant suppression of all other modes.

The injected FP laser was gain-switched at high speed by an arbitrary waveform generator (AWG), not shown in the figure, to generate optical pulses (FIG. \ref{Concept}a.iii). As each pulse was stimulated from the CW WT laser, each pulse was phase coherent with each other \cite{250390,PhysRevX.6.031044}. A high bandwidth external lithium niobate phase modulator operating at the clock-rate and driven by the AWG was then used encode the relative phase information between successive pulses (FIG. \ref{Concept}a.iv). Note that injection locking the FP laser also enhances the modulation bandwidth and reduces pulse chirp, leading to improved pulse performance and a significant advantage over pulse carving with intensity modulators \cite{8859279}.  FIG. \ref{Concept}b shows the optical path used within the transmitter. A variable optical attenuator (VOA) was used to control the power entering the circulator and hence the power of injection and a second VOA was used following the phase modulator to attenuate the output light to the desired output power before being sent to a receiver.

\section*{QKD Application}

\begin{figure}[htbp]
\centering\includegraphics[width=9cm]{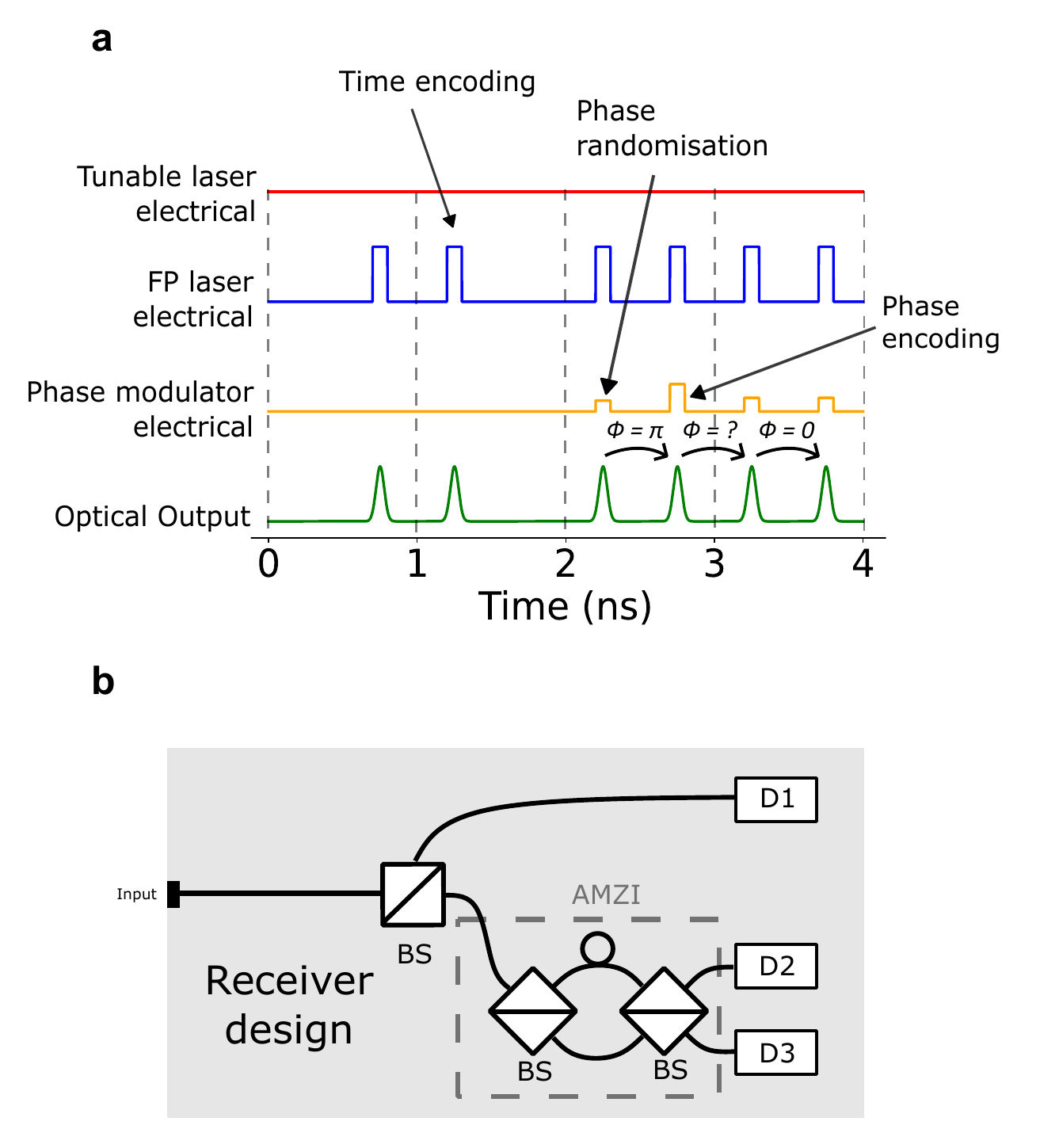}
\caption{\textbf{a} The encoding scheme used to encode binary random bits in the time and phase bases respectively and to encode discrete phase randomisation between pulse pairs. \textbf{b} The physical design of the receiver used for QKD. The light is split by a beamsplitter (BS) into two paths, either the light is directed directly to an SNSPD, D1, or it is passed through an Asymmetric Mach-Zehnder Interferometer (AMZI) where consecutive pulses are interfered and then directed to detectors D2 and D3.}
\label{Concept2}
\end{figure}

For the application of QKD, the output power of the transmitter was set such that a mean photon number of 0.4 photons per pulse was emitted. FIG. \ref{Concept2}a shows the encoding scheme used to generate information for QKD using the efficient BB84 protocol over the X and Z bases. The WT laser operates in CW with a constant DC bias, the FP laser is gain-switched providing a pulse either early $\psi = \ket{e}$ or late $\psi = \ket{l}$ with reference to a clock to encode in the Z basis or both $\psi = \frac{1}{\sqrt{2}}\big(\ket{e}+e^{i\phi}\ket{l}\big)$ to produce a superposition of early and late with a distinct phase relationship between them. In the case of a pulse being present in both time windows, the phase modulator is applied to encode the relative phase between pulses, randomly chosen from the set $\phi = \{ 0,\pi \}$ to encode in the X basis. The phase modulator is also used to produce global phase randomisation, a requirement for the BB84 protocol, where a random phase chosen from a set of 10 values spanning $0$ to $2\pi$ is applied the first pulse of every pair such that there is no cross information between symbols. Ten state discrete phase randomisation has been shown to have the equivalent security to a continuum of random phases \cite{Cao_2015}.

\begin{figure*}[t]
\centering\includegraphics[width=15cm]{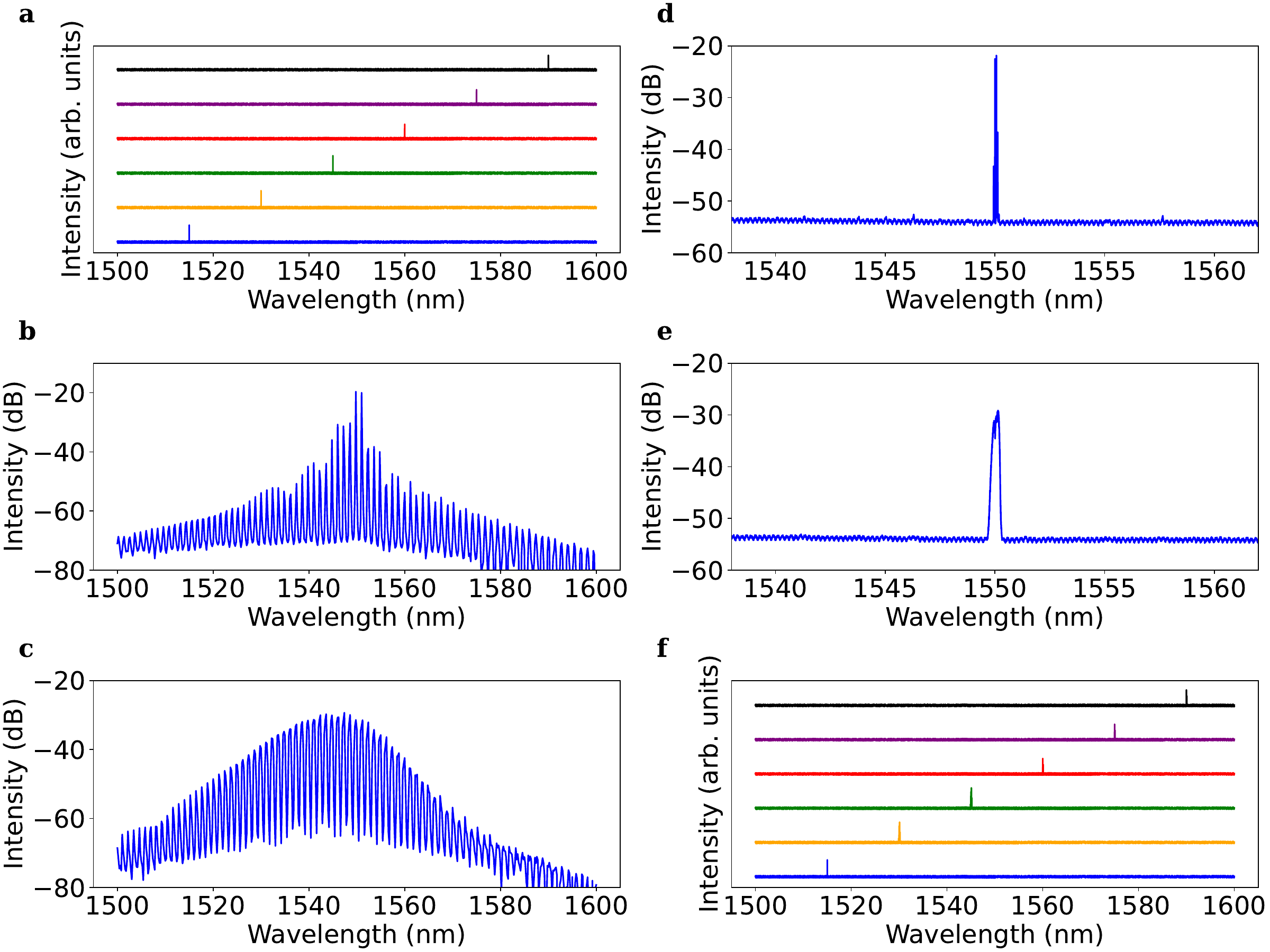}
\caption{\textbf{a} The spectral shape of the direct CW output from the WT laser operating above threshold for different set wavelengths. \textbf{b} The spectral shape of the direct CW output from the FP laser operating at 14.8 mA. A multi-modal structure is observed spanning 100 nm. \textbf{c} The spectral shape of the direct output from the FP laser operating at 14.8 mA when gain-switched with a random pattern at 2 GHz. \textbf{d} The spectral shape of the CW FP laser once injected by the CW WT laser at 1550.12 nm with maximum overlap between modes, spectrum measured at the output of the FP laser. \textbf{e} The spectral shape of the FP laser when pulsed with a random pattern at 2 GHz and injected by the CW WT laser with maximum overlap between modes. Spectrum measured at the output of the FP laser. \textbf{f} The spectral shape of the CW FP laser once injected by the WT laser at a range of different wavelengths with maximum overlap between modes. Spectrum measured at the output of the FP laser.}
\label{Wavelength}
\end{figure*}

Our transmitter design is compatible with standard QKD and novel receiver designs \cite{Beutel:22}. In this work, at the receiver, information was split via a beamsplitter (BS) into two paths where it was either directly measured at detector D1, to decode time of arrival or passed through an asymmetric Mach Zehnder interferometer (AMZI) with a delay of half the clock-rate such that adjacent pulses are interfered to detect the relative phase relationship between them (FIG. \ref{Concept2}b). Direct detection allows decoding of Z basis bits, used to generate the shared key between transmitter and receiver. Detection on either detector D2 or D3 following interference of successive pulses allows decoding of X basis bits, used to verify the security of the quantum channel and for parameter estimation.  The output of the AMZI is such that if there is no relative phase difference between successive pulse, all photons will be directed to one detector and if there is a $\pi$ phase difference between the adjacent pulses, the photons will all be directed to the other detector. The quantum bit error rate (QBER) is given as the probability that a detection event occurs in the orthogonal state with which it was encoded (i.e. detecting $\psi = \ket{l}$ when $\psi = \ket{e}$ was encoded and detecting  $\psi = \frac{1}{\sqrt{2}}\big(\ket{e}+\ket{l}\big)$ when $\psi = \frac{1}{\sqrt{2}}\big(\ket{e}-\ket{l}\big)$ was encoded). Helium cooled superconducting nanowire single photon detectors (SNSPD) operating at 2.15 K were used to detect photons within the receiver. The detectors had a detection efficiency of 33\% and dark counts of 1 Hz.

Proof of principle efficient BB84 QKD was carried out for the wavelength range 1515 nm to 1590 nm, where Z basis bits were encoded 93.75\% of the time. For each wavelength, the WT laser was tuned to the appropriate value and the FP laser TEC was tuned to maximise overlap of a single mode with the injected light. Finally the VOA that was used to attenuate to the desired output flux was adjusted such that the output was set to 0.4 photons per pulse. This was required as certain modes of the FP laser experience more gain than others and as such require higher attenuation to get to the desired flux before being input into the quantum channel. The selected wavelengths within the relevant range correspond to different channels of a Dense Wavelength-Division multiplexer (DWDM) 50 GHz International Telecommunication Union (ITU) grid. A VOA was placed between the transmitter and the receiver and QKD was performed over attenuation levels of 11-66 dB to simulate different distances of fiber between transmitter and receiver. The secure key-rate (SKR) was analysed in the asymptotic regime and was calculated through weak coherent pulse decoy state analysis \cite{Koashi,Lucamarini:13,PhysRevA.72.012326,Erven_2009}. Whilst finite key analysis was not undertaken as the QKD protocol is not the novelty of this work, the interferometer free transmitter design remained stable over many days of continuous use and therefore we conclude that a sufficient block size could be achieved to mitigate finite size effects.

\section*{Transmitter characterisation}

The WT laser can be tuned to any wavelength between 1500 nm and 1600 nm to an accuracy of 100 fm. FIG. \ref{Wavelength}a shows the direct output of the WT laser against wavelength for a range of chosen wavelengths, the linewidth is below the resolution of the spectrometer.

When operated at a low DC bias, the FP laser outputs a range of modes centered around 1550 nm and mode spacing of $\sim$ 1.25 nm. FIG. \ref{Wavelength}b shows the multi-modal behaviour of the FP laser operating at just above threshold bias (14.8 mA) at a temperature of 25 $^{\circ}$C. The central modes have a much higher gain than the edge modes. Once the FP laser is gain-switched, the pulsing changes the distribution of energy within the gain profile, giving a broader but flatter distribution, increasing the relative gain of the edge modes compared to the central ones (FIG. \ref{Wavelength}c).

Injecting the WT laser into the FP laser, the nearest mode to the chosen wavelength becomes optically injection locked to the tunable light causing one mode to dominate. Other optical modes become suppressed as the injected light dominates the available gain. This is because there are seed photons injected at this wavelength compared to the other cavity modes which are seeded by only spontaneous emission. Fine tuning of the FP laser wavelength through thermoelectric tuning allows maximisation of the injection locking and increases suppression of other modes, reducing the likelihood of mode hopping. In addition, the mode linewidth is significantly reduced due to the injection locking \cite{623252}.

FIG. \ref{Wavelength}d shows the CW behaviour of the FP laser once one mode is maximally injected with the WT laser at a wavelength of 1550.12 nm. Optimal operation was found with an optical injection power of 80 $\upmu$W  and the mode linewidth is below the resolution of the spectrometer. Once the FP laser is gain-switched to produce optical pulses, the linewidth of the selected mode broadens to a linewidth of Full-Width-Half-Maximum (FWHM) of 35 GHz (FIG. \ref{Wavelength}e).

A wide range of wavelengths can be accessed for single-mode operation by varying the wavelength of the injected mode and optimising the injection via the TEC and the input injection power. FIG. \ref{Wavelength}f shows the CW behaviour of the transmitter when set to a variety of different wavelengths between 1515 nm and 1590 nm. In all cases, the transmitter was able output a single mode with other modes suppressed by at least 30 dB. For the extreme ends of the wavelength range, a greater injection power was required to suppress competing modes due to the lower natural gain of the FP laser.

Gain-switching the injected FP laser enables the production of optical pulses. The phase of FP laser outputted pulses is set by the phase of the injected light such that all pulses are coherent with each other. Once passed through the phase modulator, an arbitrary relative phase relationship can be applied between the pulses.

A 1-dimensional Gaussian function was fitted to an 80 GHz sampling rate oscilloscope trace of the output to find a minimum pulse-width when the transmitter was optimised to output phase-controlled light. A FWHM of 69.8 ps was found at a DC bias of 14.8 mA and injection power of 80 $\upmu$W. Increasing either the injection power or the DC bias led to a slight increase in the pulse duration.

\begin{figure}[ht]
\centering\includegraphics[width=8cm]{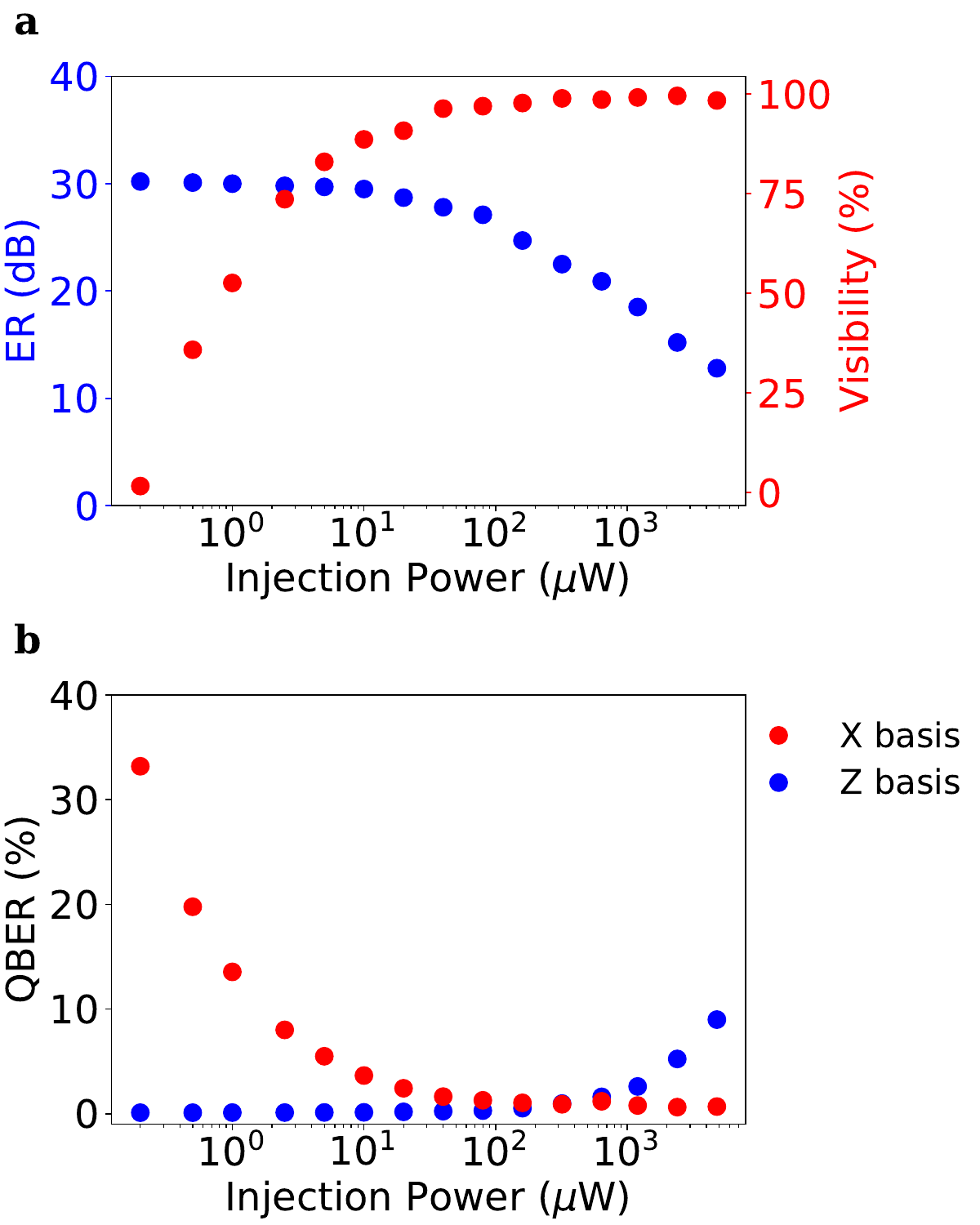}
\caption{\textbf{a} In blue, the pulse extinction ratio (ER) against injection power and in red the interference visibility of adjacent pulses against injection power. At low injection, the transmitter produces high ER pulses but exhibits multi-modal behaviour and therefore has a reduced interference visibility \textbf{b} The measured QBER against injection power. Z basis QBER is shown in blue and X basis QBER is shown in red.}
\label{ER}
\end{figure}

The transmitter performance was found to be very sensitive to the injection power of the WT laser into the FP. One metric used to define the quality of the pulses is the pulse extinction ratio (ER). The ER was measured by sending a repeating pulse pattern at a 1 GHz clock-rate from the transmitter to the single photon detectors and then taking a histogram of photon detection events against time. The ER is defined as

\begin{equation}
    ER = 10 \text{log}_{10}\bigg(\frac{N_{max}}{N_{min}}\bigg)
\end{equation}

where $N_{max}$ is the number of single photon detection events at the pulse peak and $N_{min}$ is the number of detection events at the minimum.

The best ER was found when there was no light injected into the cavity, with an ER of 30 dB. As the injection power increased, the ER reduced shown in blue in FIG. \ref{ER}a. However, with no injected light the transmitter outputs multi-modal light and hence there is no phase coherence. Therefore the interference visibility is very low. Increasing the injection power leads to greater suppression of competing modes leading to single-mode coherent behaviour and high interference visibility shown in red in FIG. \ref{ER}a. Therefore there is a trade-off between pulse extinction ratio and interference visibility and the optimal injection power is one that maximises both.

Increasing the DC bias supplied to the FP laser allows access to a greater number of modes and therefore a greater wavelength tuning range, however a greater injection power was then needed to suppress competing modes leading to a lower extinction ratio of pulses. These observations demonstrate there is great potential for tuning the exact properties of the transmitter to achieve the requirements of a particular application. We note that this optimisation could even be fully automated in a practical deployment of this design, for example, by leveraging recent advances in machine intelligence as shown in Ref. \cite{PhysRevApplied.18.034087}.

\section*{Wavelength tunable QKD}

Once a single mode was selected, QKD could be performed. In the context of QKD, a lower ER leads to an increase in the Z basis QBER shown in blue in FIG. \ref{ER}b. This is because there is a higher proportion of counts in the opposing time window to that in which the information was encoded. In addition, a low interference visibility leads to an increase in the X basis QBER shown in red in FIG. \ref{ER}b as some photons will be directed to the incorrect detector within the receiver following interference. An injection power of between 80 $\upmu$W and 160 $\upmu$W is able to minimise the total QBER.

\begin{figure}[ht]
\centering\includegraphics[width=8cm]{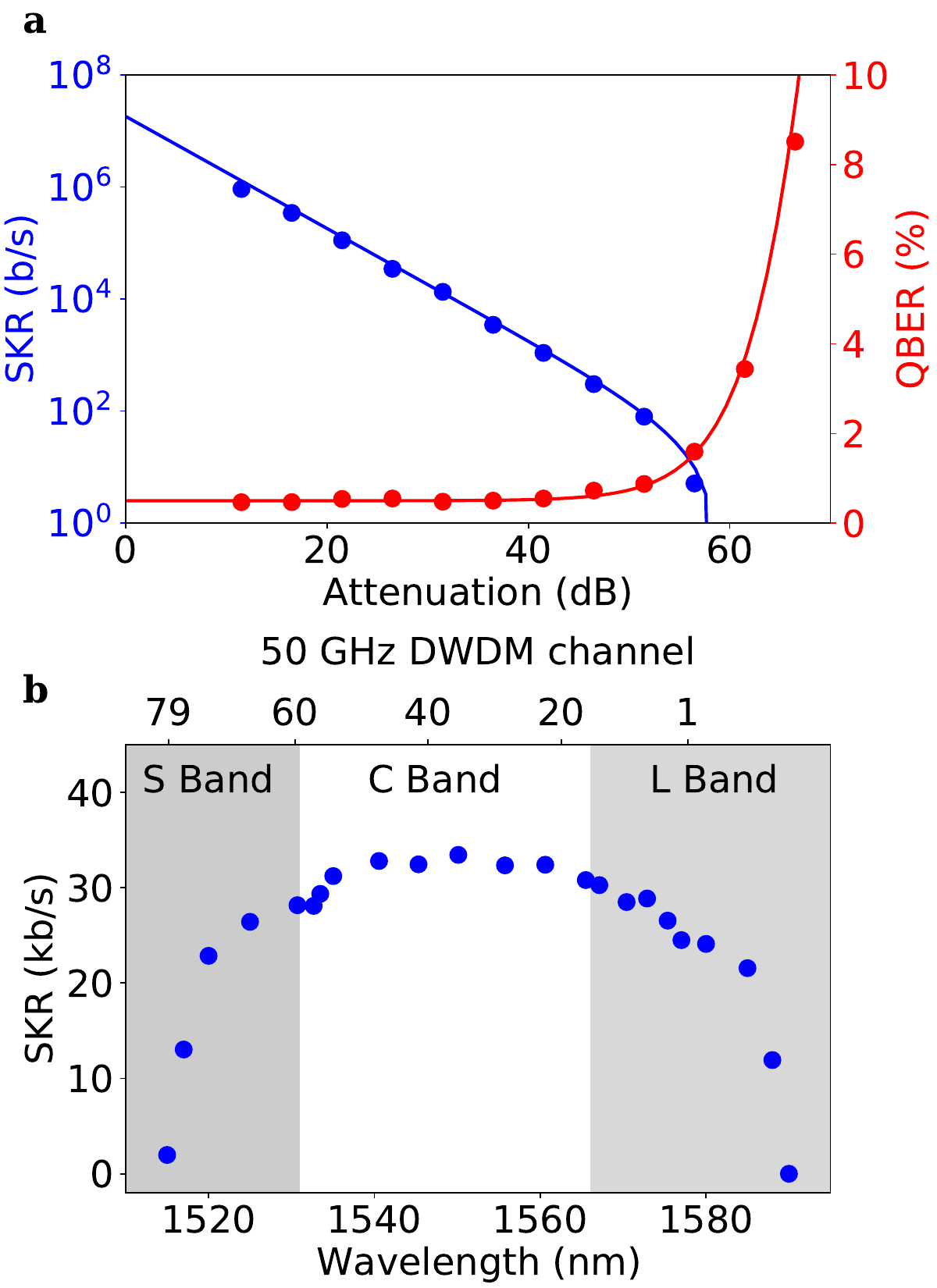}
\caption{\textbf{a} The secure key-rate (blue) and the QBER (red) vs attenuation for the transmitter operating at 1550.12 nm. The dots show the measured values and the lines show simulated values. \textbf{b} The secure key-rate vs wavelength at an attenuation of 26.5 dB. The telecommunications C band is highlighted in white whilst the S and L bands are highlighted in grey. Wavelengths within range correspond to channels of a 50 GHz DWDM ITU grid.}
\label{Keyrate}
\end{figure}

FIG. \ref{Keyrate}a shows the measured secure key-rate and QBER of the transmitter against channel attenuation once optimised to operate at a wavelength of 1550.12 nm. This is plotted alongside simulations calculated by inputting equipment parameters into the SKR formula. A maximum SKR of 935 kb/s was found at 11.5 dB attenuation giving a simulated 17.7 Mb/s key-rate at 0 dB transmission. Data-points were not collected at lower attenuations due to the count-rate nearing the saturation point of the SNSPD detectors.

The transmitter allows QKD to be performed at a variety of wavelengths. FIG. \ref{Keyrate}b shows how the SKR that can be achieved changes against operating wavelength when QKD was performed over a fixed attenuation of 26.5 dB. The distribution of SKR against wavelength is relatively flat, all key-rates measured within the range 1520 nm to 1585 nm are within a factor of 1.5 of each other. Across the telecommunications C band, highlighted with a white background, the key-rate only drops to 85\% of its maximum measured value meaning that usable QKD keys can be generated at any wavelength within the C band.

The best SKR was measured at 1550.12 nm. This is to be expected as the FP laser is centered around this wavelength; this mode has a greater natural gain profile than an edge mode, therefore a lower injection power is needed for it to dominate the gain profile of the laser. In addition, the circuit components such as the AMZI in the receiver, the attenuators used to simulate the quantum channel and the detectors, were designed to be operated at that wavelength and are likely to have slightly reduced performance at other wavelengths. As the wavelength moves away from 1550.12 nm, there is a small roll-off in the optimised key-rate. This is due to the reduced gain profile of the FP laser at these wavelengths meaning that there is greater mode competition from non-injected modes and as such, a higher injection power was needed to cause a single mode to dominate. The increased injection power leads to a slightly higher QBER due to a reduced pulse extinction ratio and hence a reduced key rate. At 1515 nm and 1590 nm, the key-rate drops significantly as the inherent gain profile of the FP laser is very weak and even with a very high injection power, there is significant mode competition, leading to poor interference visibility and a high X basis QBER. The drop in key-rate is also partially explained by the detectors, which were designed for use at 1550 nm, hence there is a slight reduction in efficiency as the wavelength is moved further from this wavelength. 

\section*{Discussion}

A wavelength tunable optical transmitter that is able to produce phase-controlled optical pulses has applications spanning classical and quantum communication networks. The flexibility that has been gained by implementing wavelength tunability is particularly applicable to next generation communication networks such as dynamically reconfigurable software-defined networks. For the application of QKD, being able to switch between wavelengths could enable the same network to be used by multiple users with different wavelength requirements in a time-multiplexed fashion. This could potentially reduce the complexity of a large passive optical network making large scale QKD significantly easier to implement and cheaper. In addition, the components required to create this wavelength tunable transmitter are off-the-shelf, no custom built modules are needed, therefore to upgrade from a standard QKD transmitter that is commercially available today to a wavelength tunable transmitter is readily available and low-cost. The design is also significantly lower cost than the alternative design of having a separate fixed wavelength system at each ITU channel with an optical switch to transfer between them. In addition, in contrast to many available QKD transmitters, this transmitter design does not include an interferometer and is therefore more stable over long time-periods. Indeed, the system could run continuously over a period of many days, keeping the QBER low without need for realignment. 

The gain-switching design is preferential over placing external intensity modulators after a CW source as the achievable ER is significantly greater than can be achieved by an intensity modulator and is cheaper and more compact than placing a chain of intensity modulators in the beam-path. The key parameter when gain-switching lasers is the direct modulation bandwidth which can be approximated by

\begin{equation}
    f_{3 dB} \approx \frac{3}{4\pi^2 q} \frac{\Gamma v_g \sigma_g}{V}(I_b - I_{th})
\end{equation}

where $q$ is the elementary charge, $\Gamma$ is the confinement factor, $v_g$ is the group velocity of light in the cavity, $\sigma_g$ is the differential gain, $V$ is the active volume, $I_b$ is the bias current and $I_{th}$ is the threshold current \cite{8859279}. Here, the bandwidth is proportional to the confinement factor of the laser cavity and inversely proportional to the active volume, hence why direct modulation of a WT laser with a tunable cavity does not allow a high qubit encoding rate. FP lasers, on the other hand can be modulated at high speed as they have no tunable elements and are the most simplistic laser design. The combination of the WT laser operating in CW and injecting a single mode of the FP laser therefore allows a wavelength tunable single mode that can be modulated at high speed.

An alternative approach to generate phase modulation is through direct modulation of the injected light which changes the refractive index, and hence path length of the laser cavity mode, hence creating a phase shift \cite{yuan2016directly}. Direct modulation of the wavelength tunable laser did allow for some phase control, however, the modulation bandwidth of an external cavity laser is too low for reliable phase control at GHz speeds. In addition, the phase modulation that could be achieved through direct modulation of the external cavity laser was not the full 2 $\pi$ and hence the decision to use a phase modulator was taken.

The transmitter could also be used to conduct QKD using other protocols such as the Coherent-One-Way (COW) protocol \cite{DeMarco:21,https://doi.org/10.1002/lpor.201700067} or the Measurement Device Independent (MDI) and Twin-Field (TF) protocols \cite{RobMDI,MariellaTF} by changing the modulation pattern applied to the phase modulator and the FP laser.

Further wavelength bands could be reached by changing the semiconductor material in the laser sources. The cause in the drop-off in achievable key-rate occured due to the reduced gain of the FP laser at the extreme ends of the wavelength range. If a different FP laser with a different gain profile was used, this would lead to a different range in tunability. Through this method, tunable pulsed laser sources could be produced centered in different telecommunication bands such as the S, L or O Bands. 

A potential avenue to be explored further is to directly modulate the WT laser in order to encode relative phase information between pulses rather than using an external LiNbO$_{3}$ phase modulator. Here, a small change in the driving current supplied to the WT laser could be applied, causing a phase shift - a similar idea with fixed wavelength sources has been shown to work already~\cite{chipmodfree}.

\section*{Conclusion}

Here we have presented a wavelength tunable optical transmitter capable of producing phase-controlled pulses at a rate of 2 GHz and used it to implement wavelength tunable QKD at a symbol clock-rate of 1 GHz. We have demonstrated that high secure key rates can be achieved with this transmitter over a wavelength range of 65 nm. The tunability of the quantum channel allows for flexibility in the delivery of QKD enabling the wavelength utilized to be optimised to the requirements of a user. This opens the gateway to wavelength multiplexing multiple quantum signals upon the same fiber without each multiplexed system requiring bespoke hardware or chains of intensity modulator and opens the way to dynamically reconfigurable QKD, an emerging and exciting prospect for many metropolitan networks where user needs change regularly.

\section*{Acknowledgments}
This research includes the results of research and development project of ICT priority technology (JPMI00316) `Research and Development for Building a Global Quantum Cryptography Communication Network' by Toshiba and MIC.

\end{document}